\input epsf
\documentstyle[preprint,aps]{revtex}
\begin{document}
\draft
\tighten
\title{Enhancement  of disoriented chiral condensate domains with
friction}
\author{\bf A. K. Chaudhuri \cite{byline}}
\address{ Variable Energy Cyclotron Centre\\
1/AF,Bidhan Nagar, Calcutta - 700 064\\}
%\date{April 5, 1999}
%Submitted to Phys. Rev. C on August, 1998
\maketitle
\begin{abstract}
We  investigate  the  effect  of  friction on domain formation in
disoriented chiral condensate.  Including  a  friction  term,  we
solve the equation of motion of the linear sigma model fields, in
the Hartree approximation. With boost-invariance and cylinderical
symmetry,  irrespective  of  friction, {\em on average}, we donot
find any  indication  of  domain  like  formation  with  quenched
initial condition. However, with or without friction, some events
can  be  found  with large instabilities, indicating possible DCC
domain formation in those events. With friction time scale during
which  instabilities  grows  increases.   Correspondingly,   with
friction,  it  is  possible to obtain large sized domains in some
particular events.
\end{abstract}


The possibility of forming disoriented chiral condensate (DCC) in
relativistic  heavy  ion  collisions  has  generated considerable
research activities in recent years. The idea was first  proposed
by Rajagopal and Wilczek \cite{ra93,ra93a,ra95,wi93}. They argued
that  for  a  second  order  chiral  phase transition, the chiral
condensate   can   become   temporarily   disoriented   in    the
nonequilibrium conditions encountered in heavy ion collisions. As
the  temperature drops below $T_c$, the chiral symmetry begins to
break  by  developing  domains  in  which  the  chiral  field  is
misaligned  from its true vacuum value. The misaligned condensate
has the same quark content and quantum numbers as  do  pions  and
essentially  constitute  a  classical pion field. The system will
finally relaxes to the true vacuum and in the  process  can  emit
coherent  pions.  Since the disoriented domains have well defined
isospin orientation,  the  associated  pions  can  exhibit  novel
centauro-like  \cite{la80,ar83,al86,re88} fluctuations of neutral
and charged pions \cite{an89,an89a,bj93,bl92}.

Most dynamical studies of DCC have been based on the linear sigma
model,  in  which  the chiral degrees of freedom are described by
the real O(4)  field  $\Phi=(\sigma,\roarrow{\Pi})$,  having  the
equation of motion,

\begin{equation}
[\Box + \lambda (\Phi^2-v^2)]\Phi=H n_\sigma
\label{1}
\end{equation}

The  parameters  of the model can be fixed by specifying the pion
decay constant, $f_\pi$=92 MeV and the meson masses,  $m_\pi$=135
MeV       and       $m_\sigma$=600      MeV,      leading      to
$\lambda=(m_\sigma^2-m_\pi^2)/2f_\pi^2$=20.14                 and
$v=[(m_\sigma^2-3            m_\pi^2)/(m_\sigma^2-m_\pi^2)]^{1/2}
f_\pi$=86.71  MeV  and  $H=(120.55  MeV)^3$  \cite{ra96}.  It  is
apparent  from  eq.\ref{1}  that  the  vacuum  is  aligned in the
$\sigma$  direction  $\Phi_{vac}=(f_\pi,\bf{0})$   and   at   low
temperature  the  fluctuations represent nearly free $\sigma$ and
$\pi$ mesons. At very high temperature well above $v$, the  field
fluctuations are centered near zero and approximate O(4) symmetry
prevails.

It is instructive to decompose the chiral field,

\begin{equation}
\Phi(r,t)=<\phi(r,t)>+\delta\phi(r,t) \label{2}
\end{equation}

\noindent  where $<\phi> $is the mean field and $\delta \phi$ are
the  semiclassical  fluctuations  around  $<\phi>$  and  can   be
identified  with quasi-particle excitations. Using eq.\ref{2} and
taking the average of eq.\ref{1}, the equation of motion for  the
mean fields in the  Hartree  approximation  can  be  obtained  as
\cite{ra96,as94},

\begin{equation}
\frac{\partial^2 <\phi>}{\partial t^2} -\nabla^2 <\phi>
= \lambda(v^2-<\phi>^2-
3<\delta  \phi^2_\|>-<\delta  \phi^2_\bot>)  <\phi>  +H  n_\sigma
\label{3}
\end{equation}

\noindent  where $<\phi>^2=<\phi_i><\phi_i>$, $\delta \phi_\|$ is
the component of the fluctuation parallel to $<\phi>$ and $\delta
\phi_\bot$ is the orthogonal component. This equation imply  that
the  motion  of  the  mean  field  is determined by the effective
potential,

\begin{equation}
V(<\phi>)=\frac{\lambda}{4}
(<\phi>^2+ 3<\delta \phi^2_\|>+<\delta \phi^2_\bot> -v^2)
\label{4}
\end{equation}

\noindent  which clearly differs from the zero temperature one in
presence of fluctuations. By  varying  the  fluctuations,  chiral
symmetry  can  be  restored  or  spontaneously broken. It is also
evident that the evolution of the mean field  critically  depends
on  the initial values of the fluctuations. When $\delta^2 \equiv
(3<\delta \phi^2_\|>-<\delta \phi^2_\bot>)/6$ is large enough the
chiral symmetry is approximately (as H$\neq$ 0) restored. If  the
explicit  chiral  symmetry  breaking term is neglected, the phase
transition   takes   place   at   the    critical    fluctuations
$\delta^2_c\equiv   v^2/6$.  For  $\delta^2  <  \delta^2_c$,  the
effective    potential    takes    its    minimum    value     at
$<\phi>=(\sigma_e,0)$,  where  $\sigma_e$  depends on $\delta^2$.
When the mean fields are displaced from this equilibrium point to
the central  lump  of  the  Mexican  hat  ($<\phi>\sim  0$),  the
effective mass square

\begin{equation}
m_{eff}^2=\lambda(v^2-<\phi>^2-3<\delta \phi^2_\|>-<\delta
\phi^2_\bot>) \label{5}
\end{equation}

\noindent will become negative and DCC can form. Since the domain
size  is  directly  related  to  the time scale, during which the
effective mass remains  negative,  it  strongly  depends  on  the
initial  condition  of  the system. By varying the $<\phi_i>$ and
$\delta^2$, quench or annelaing like  initial  condition  can  be
obtained  \cite{as94}.  In  an  important  paper  Asakawa  et  al
\cite{as94}   studied   eq.\ref{3}   with    initial    condition
corresponding to quench and annealing. They found that domains of
disoriented  chiral  condensate  with  4-5  fm  in  size can from
through a quench. Annealing on the otherhand,  leads  to  smaller
sized domains.

In  the  present  paper,  we  solve  eq.\ref{3}  with  a diffrent
motivation. Our interest is to investigate the effect of friction
on DCC domain formation. Dissipative effects like  friction  damp
the motion of the fields, inhibiting large oscillations. Naively,
one would expect reduction of DCC domain formation with friction.
Recently,  Biro  and  Greiner  \cite{bi97},  using  the  Langevin
equation for the linear sigma model, investigated  the  interplay
of  friction  and  white  noise  on  the  evolution  of the order
parameter. While noise greatly diminishes the possibilty  of  DCC
domain  formation, in some orbits, large instabilities can result,
producing DCC  domains.  We  have  also  studied  the  effect  of
friction  on  DCC  domain  formation  using the Langevin equation
\cite{ch99}. There we found that for one-dimensional expansion on
average large DCC domain can not  be  formed.  However,  in  some
particular  orbit  large instabilities can occur. This possibilty
also reduces with  introduction  of  friction.  However,  if  the
friction is large, the system may be overdamped and then there is
a  possibilty  of  DCC domain like formation. Present paper is an
extension of the above study, the spatial part,  which  had  been
integrated out in our earlier analysis is being studied here.

Appropriate  coordinates  for heavy ion collisions are the proper
time  ($\tau$)  and  the  rapidity  (Y).  For   a   d-dimensional
expansion,   the   change   can  be  effected  by  the  following
replacement,

\begin{equation}
\frac{\partial^2}{\partial t^2}- \frac{\partial^2}{\partial z^2}
\rightarrow
\frac{d}{\tau} \frac{\partial}{\partial \tau} \tau
\frac{\partial}{\partial \tau} -
\frac{1}{\tau^2} \frac{\partial^2}{\partial Y^2} \label{6}
\end{equation}

It can be seen from the above equation that with the introduction
of proper time and rapidity, a dissipative term comes into effect
in the equation of motion. To illustrate the role of friction, we
further introduce a dissipative term ($\eta$) in the equation. To
simplify our calculation, we  assume boost-invariance as well
as cylindirical symmetry in the system. With boost-invariance and
cylindirical symmetry, eq.\ref{3} can be written as,

\begin{equation}
\frac{\partial^2 <\phi>}{\partial \tau^2} +(\frac{d}{\tau}+\eta)
\frac{\partial <\phi>}{\partial \tau} =
\frac{\partial^2 <\phi>}{\partial r^2} +\frac{1}{r}
\frac{\partial <\phi>}{\partial r} +
 \lambda[v^2-<\phi>^2-
T^2/2] <\phi> +H n_\sigma
\label{7}
\end{equation}

\noindent  where  $\eta$  is the friction coefficient and we have
replaced the fluctuation terms by their counterpart ($T^2/2$)  in
the   finite   temperature  field  theory  \cite{ga94,ra97}.  The
equation of motion of  fields  then  depend  sensitively  on  the
initial  temperature  of  the system. In the following, we assume
initial              temperature              to               be
$T_c=\sqrt{2f_\pi^2-2m_\pi^2/\lambda}$=123  MeV  at  the  initial
time $\tau_i$= 1 fm. The cooling of the system  is  described  by
the equation,

\begin{equation}
\frac{\dot{T}}{T} + \frac{d}{3\tau}=0 \label{8}
\end{equation}

In the weak coupling limit, the friction $\eta$ is related to the
on-shell plasmon damping rate, $\eta \equiv 2\gamma_{pl}$. In the
standard  $\phi^4$  theory,  the  plasmon  damping  rate  can  be
calculated   for   the   $\sigma$   and    the    $\Pi$    fields
\cite{gr97,pa92}.  Assuming  that  the meson masses are the same,
the friction coefficient can be obtained as \cite{bi97},

\begin{equation}
\eta=2\gamma_{pl}=\frac{9}{16\pi^3}\lambda^2        \frac{T^2}{m}
f_{Sp}(1-e^{-m/T}) \label{9}
\end{equation}

\noindent  where  $f_{Sp}=-\int^x_1  dt \frac{\ln t}{t-1}$ is the
Spence function. At  $T=T_c=\sqrt{2f^2_\pi-2m^2_\pi/\lambda}$=123
MeV,  and if $m/T \simeq 1$, the friction $\eta= 2.2 fm^{-1}$. In
order to have a  comparison,  we  also  consider  scenarios  with
$\eta$=0   and   $\eta$=0.5   $fm^{-1}$.   We  also  neglect  the
temperature dependence of the friction coefficient.

We  solve  the set of partial differential equations \ref{7} with
the  quenched  initial  condition,  for  two  dimensional   (d=2)
expansion  using  the standard Leap frog method. Evolution of the
fields upto $\tau$=7 fm  are  followed.  As  argued  before,  DCC
formation  depend critically on the initial condition. To reflect
the uncertainty in the initial condition, the initial fields  are
randomly  distributed  to  a  Gaussian  form  with  the following
parameters \cite{as94},

\begin{eqnarray}
<\sigma>=&&(1-f(r))f_\pi \\
<\pi_i>=&&0 \\
<\sigma^2>-<\sigma>^2 = <\pi_i^2>-<\pi_i>^2=   && v^2/6 f(r)\\
%<\pi_i^2>-<\pi_i>^2=&& v^2/6 f(r)\\
< \dot{\sigma}>=&& <\dot{\pi_i}>=0\\
<\dot{\sigma}^2>=<\dot{\pi}>^2=&&4 v^2/6
\end{eqnarray}

The interpolation function

\begin{equation}
f(r)=[1+exp(r-r_0)/\Gamma)]^{-1}
\end{equation}

\noindent separates the central region where  the  initial  field
configuration  is  different from their vaccum expectation value.
We use $r_0$=5 fm and $\Gamma$=0.5 fm.

The  equation of motion was solved for 500 trajectories. For each
trajectories, at each space-time we compute  the  effective  mass
$m^2_{eff}$.  The phenomenon of long wavelength DCC amplification
will occur whenever the effective mass squared  is  negative.  To
single out the trajectories for which maximum instabilities occur
for each trajectories, we calculate the following quantity,

\begin{equation}
G=\int  |m_{eff}| \Theta(-m_{eff}^2) d\tau dr
\end{equation}

This can be a measure of instability in a particlar evolution. We
call  it amplification factor. This is an important parameter, as
it directly relates to the size of DCC domains. In
\begin{figure}
\centering
\epsfysize=5 cm
\epsffile[19 409 534 709]{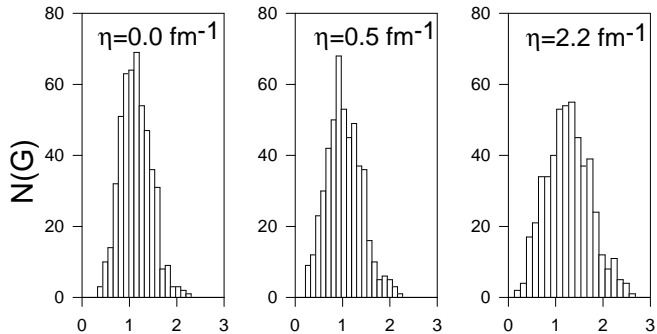}
%\vspace{.6cm}
\caption{Distribution of the amplification factor G for the three
scenarios  (i)  $\eta$=0,  (ii)  $\eta$=0.5  $fm^{-1}$  and (iii)
$\eta$=2.2 $fm^{-1}$. }
\end{figure}
\noindent fig.1, we have
shown  the distribution of the amplification factor G for the 500
trajectories. Three scenarios; (i) friction free case ($\eta$=0),
(ii)  small  friction  ($\eta$=0.5  $fm^{-1}$)  and  (iii)  large
friction  ($\eta$=2.2  $fm^{-1}$)  are  shown. Several intersting
features emerges. Maximum amplification factor is increased  with
friction.  However  the  increase  is  small ($G_{max}$= 2.19 for
$\eta$=0 and $G_{max}$= 2.55 for $\eta$=2.2  $fm^{-1}$).  Minimum
amplification  factor  consequently decreases with friction. As a
result, distribution  is  broadened  with  friction.  The  result
suggest  that  with friction, while number of orbits with minimum
instabilities are  increased,  some  orbits  can  be  found  with
enhanced  instabilities.  Thus  with  friction  possiblity of DCC
domain like formation  is  increased  in  some  orbits.  This  is
contrary  to the naive expectation that friction will inhibit DCC
like formation. However, this is in agreement  with  our  earlier
study  that  with friction some trajectories can show appreciable
instabilities \cite{ch99}.

In  fig.2,  we  have  shown  the  contour  plot of $sgn(m_{eff})$
obtained by averaging the fields over the 500 events, as obtained
in the three scenarios, (i)$\eta$=0, (ii)$\eta$=0.5 $fm^{-1}$ and
(iii)$\eta$=2.2  $fm^{-1}$.  Friction  lowers  the  magnitude  of
$m^2_{eff}$  is  evident  from  the  figure. However, $m^2_{eff}$
remain positive throughout the space-time region in all the three
scenarios.  The  result  indicate  that  on  averaging,  quenched
initial condition do not lead to DCC domain formation. Though not
shown,  average  fields  are  also found to be order of magnitude
lower than obtained in an individual event. The  result  suggests
that on average DCC like phenomena will not be obserevd.

The  behavior  is changed in individual events. In fig.3, we have
shown the contour plot of $sgn(m_{eff})$ for the  most  unstable
orbit i.e. for which the amplification factor is the maximum. The
three  cases  are  labeled  appropriately.  For the friction free
case, the effective mass remains  negative  over  a  considerable
space-time.  Thus  from  from  $\tau$=2-4  fm over a considerable
radial range (2-4 fm), $m^2_{eff}$ is negative and one may expect
DCC like formation  correspondingly.  With  small  friction,  the
pattern remains essentially the same, but space-time region where
$m_{eff}$  remains  negative  is  increased.  Entirely  different
behavior results when friction is high ($\eta$=2.2 $fm^{-1}$). In
some radial ranges,  soon  after  the  start  of  the  evolution,
$m^2_{eff}$  becomes negative and remains negative throughout the
evolution. This  is  an  indiaction  of  overdamped  system.  The
viscous  drag  is  large  and once the system enters the unstable
region viscous drag  forces  it  remain  there.  This  result  is
interesting,  as  it  suggest  possiblity of forming large domain
like structure with high viscous force in the system.

In  fig.4,  the  contour  plot  of the $\pi_1$ field for the most
unstable orbit are shown. The fields have been scaled by  $f_\pi$.
For  the  friction  free  case  several regions can be identified
where growth  of  negative  $\pi_1$  is  considerable.  Thus  for
example around r=6 fm, negative $\pi_1$ grows for about 4-5 fm of
(proper)  time.  The  results  for  small friction case is nearly
similar. We observe that for friction  free  and  small  friction
case, DCC domain like structure can grow for about 4 fm. However,
the  radial  extension  is not large, it is about 1 fm. For large
friction  case  ($\eta$=2.2  $fm^{-1}$,  the  pattern  is   again
different  and  reminiscent  of  the $m^{eff}$. We find strips of
positive and negative $\pi_1$ again of radial extension of  about
1  fm  starts  growing  just  after  the  start of the evolution.
Evidently, with large friction the domains can grow for a  longer
duration.

\begin{figure}
\centering
\epsfysize=16 cm
%\vspace{-.6cm}
\epsffile[14 13 581 829]{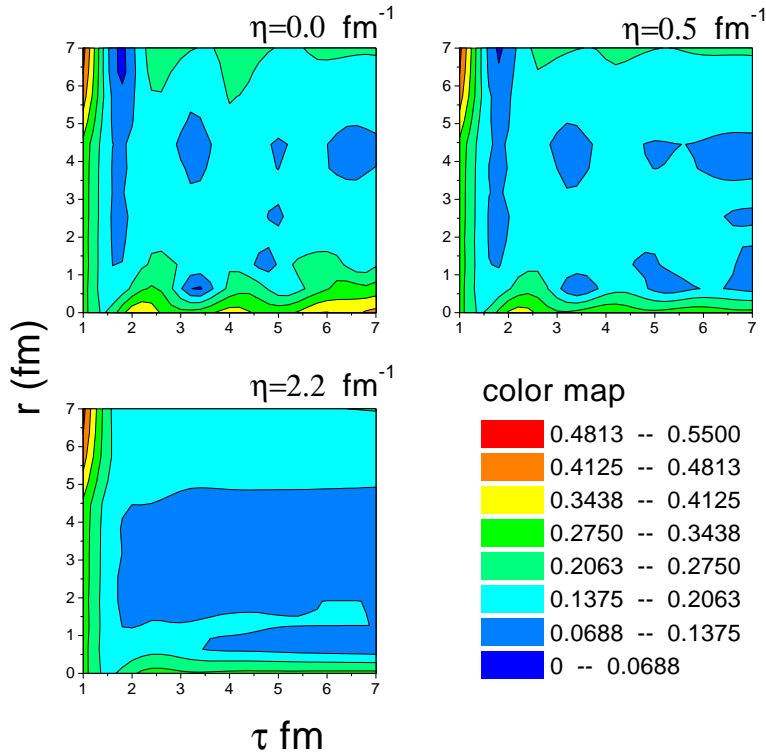}
\vspace{-2cm}
\caption{Contour  plot of evolution of $sgn(m_{eff})$ obtained by
averaging  the  fields  over  the  500  events  for  (i)$\eta$=0,
(ii)$\eta$=0.5 $fm^{-1}$ and (iii)$\eta$=2.2 $fm^{-1}$.}
\end{figure}
\begin{figure}
\centering
\epsfysize=16 cm
%\vspace{-.6cm}
\epsffile[14 13 581 829]{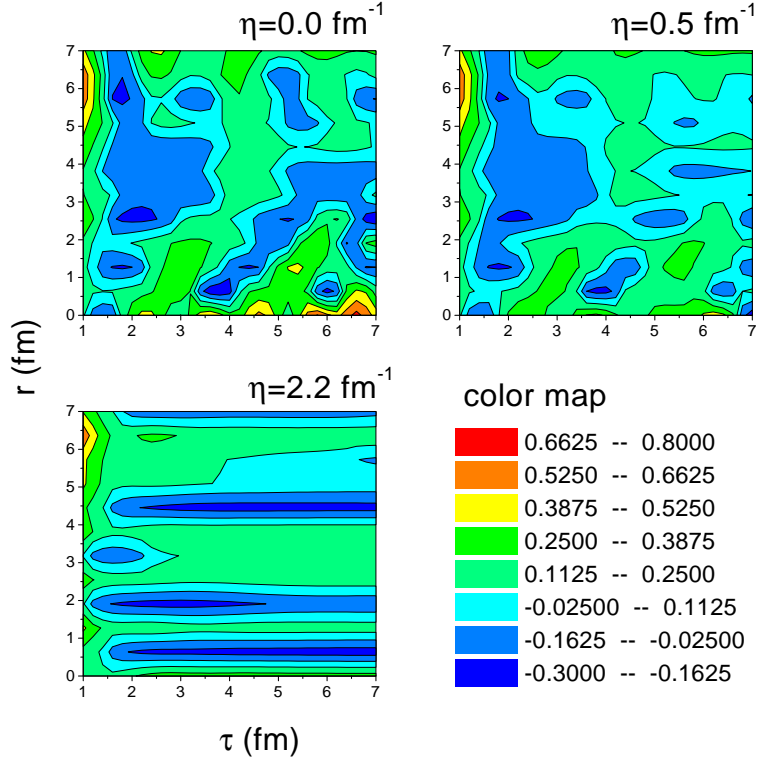}
\vspace{-2cm}
\caption{Contour  plot  of  evolution  of  $sgn(m_{eff})$ for the
event for which the amplification  factor  is  (G)  the  maximum.
Three   scenarios   (i)$\eta$=0,   (ii)$\eta$=0.5  $fm^{-1}$  and
(iii)$\eta$=2.2 $fm^{-1}$ are shown.}
\end{figure}
Before  we  summarise,  it  is  important  to  note  one apparent
contradiction we observe presently. For the  friction  free  and
small friction case domain like structure of $\pi_1$ field do not
exactly  follow  the  pattern  of  $sgn(m_{eff})$.  Thus in those
space-time region where $m^2_{eff}$ is most  negative,  we  donot
find  large amplitude of $\pi_1$ field. Though not shown, we have
checked that this is true for $\pi_2$ and  $\pi_3$  fields  also.
However  in  the  large friction scenario, $\pi_1$ follows nearly
the pattern of  $m^2_{eff}$.  The  reason  may  be  the  symmetry
breaking  term. The assumption that phenomenon of long wavelength
oscillation occur whenever  $m^2_{eff}$  is  negative  implicitly
neglects the symmtery breaking term. However, for small friction,
symmetry   breaking  term  may  not  be  negligible.  Then,  long
wavelength oscillations will not follow exactly $sgn(m_{eff})$.
\begin{figure}
\centering
\epsfysize=16 cm
%\vspace{-.6cm}
\epsffile[14 13 581 829]{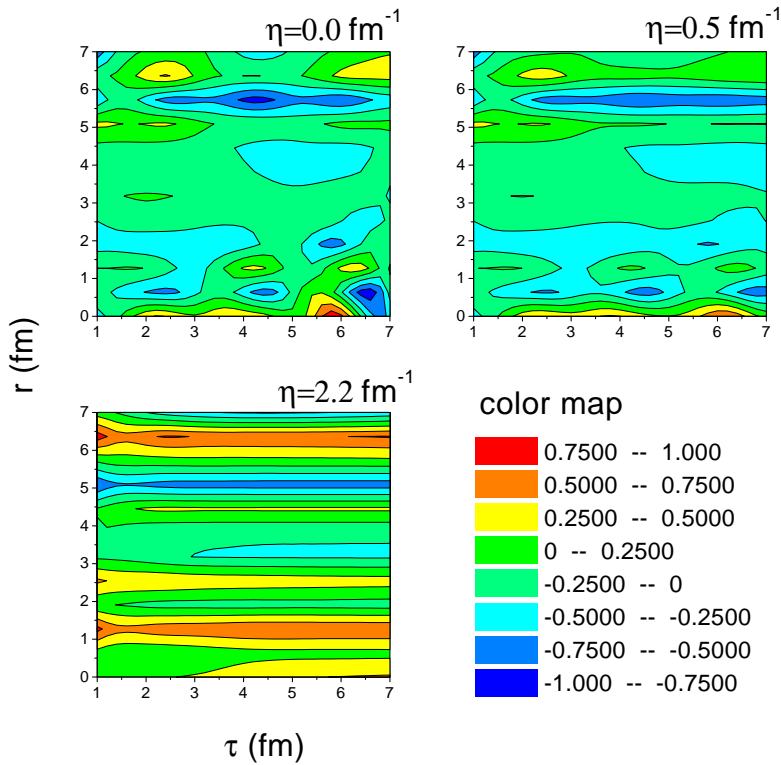}
\vspace{-2cm}
\caption{Contour plot of evolution of $\pi_1/f_\pi$ for the event
for  which  the  amplification  factor  (G) is the maximum. Three
scenarios    (i)$\eta$=0,    (ii)$\eta$=0.5     $fm^{-1}$     and
(iii)$\eta$=2.2 $fm^{-1}$ are shown.}
\end{figure}

To  summarise, we have investigated the effect of friction on the
possible DCC domain formation. In the equation of mation for  the
linear  sigma  model fields, we include a friction term and solve
it assuming boost invariance and cylinderical  symmetry.  Initial
field  configuration  was  assumed  to  be  Gaussian  random with
quenched condition ($<\phi>=\dot{<\phi>}=0$). At each  space-time
$m^2_{eff}$   was   calculated.  Phenomenon  of  long  wavelength
amplification  occurs  when  it  becomes  negative.   $m^2_{eff}$
obtained  by  averaging the field over 500 events remain positive
through out the space-time region investigated,  indicating  that
on  average,  DCC domain like formation is not possible. However,
instabilities can occur  in  some  particular  events.  In  those
events,  where the instabilites are the largest, we find evidence
of domain like structure formation. For friction free  and  small
friction,  domains  can  grow  for  about  3-4 fm. But with large
friction, the system  become  overdamped  and  instabilities  can
continue for longer duration.


\begin{references}
\bibitem[*]{byline}e-mail address:akc@veccal.ernet.in
\bibitem{ra93}   K.   Rajagopal   and  F.  Wilczek,  Nucl.  Phys.
B{\bf379}, 395(1993).
\bibitem{ra93a}   K.   Rajagopal   and   F.  Wilczek,Nucl.  Phys.
B{\bf404}, 577(1993).
\bibitem{ra95}  K. Rajagopal,in Quark-Gluon Plasma 2, ed. R. Hwa,
World Scientific (1995).
\bibitem{wi93}F.  Wilczek,  hep-ph/9308341.  Phys.  Reports {\bf
65},151(1980).
\bibitem{la80}C. M. G. Lattes, Y. Fujimoto and S. Hasegawa, Phys.
Reports, {\bf 154},247(1980).
\bibitem{ar83}  G.  Arnison  et  al,  Phys. Lett. B {\bf 122},189
(1983).
\bibitem{al86}G.  J.  Alner,  Phys.  Lett. B {\bf 180}415,(1996);
Phys. Reports {\bf 154},274 (1987). \bibitem{re88} J. R.  Ren  et
al, Phys. Rev. D {\bf 38},1517 (1988).
\bibitem{an89}  A.  A.  Anselm  and  M. G. Ryskin , Phys. Lett. B
{\bf266},482 (1989).
\bibitem{an89a} A. A. Aneslm, Phys. Lett. B{\bf 217},  169(1989).
\bibitem{bj93}J.  D.  Bjorken,  K. L. Kowalski and C. C. Taylor ,
SLAC-PUB-6109 (1993), K.L. Kowalski  K.  L.  and  Taylor  C.  C.,
CWRUTH-92-6 (1992),hep-ph/9211282.
\bibitem{bl92}  J. P. Blaizot ,A. Krzywicki , Phys. Rev.D{\bf46},
246 (1992).
\bibitem{ra96}J. Randrup, Report LBL-38125(1996), hep-ph/9602343.
\bibitem{as94} M. Asakawa, Z. Huang and X.N. Wang, hep-ph/9408299.
\bibitem{bi97}T.  S.  Biro  and C. Greiner, Phys. Rev. Lett. {\bf
79}, 3138(1997).
\bibitem{ch99}  A.  K.  Chaudhuri, nucl-th/9809018, Phys. Rev. D.
(in press).
\bibitem{ga94}  S. Gavin and B. Muller , Phys. Lett.B{\bf329},486
(1994).
\bibitem{ra97}J. Randrup, Phys. Rev. D {\bf 55},1188(1997).
\bibitem{gr97}  C.  Greiner  and B. Muller, Phys. Rev. D{\bf 45},
1026 (1997).
\bibitem{pa92}R. P. Parwani, Phys. Rev. D 45,4695(1992); S. Jeon,
Phys. Rev. D.{\bf 52},3591(1995).
\end{references}
\end{document}